\documentclass[11pt,twoside]{article}

\usepackage{asp2006}
\usepackage{epsf}
\usepackage{psfig}
\usepackage{lscape}

\begin{document}

\title{VSOP Observations of the Blazar S5 2007+77}
\author{K. \'E. Gab\'anyi,\altaffilmark{1,2,3} T. P. Krichbaum\altaffilmark{4}, A. Kraus\altaffilmark{4}, A. Witzel\altaffilmark{4} and J.A. Zensus\altaffilmark{4}}
\affil{$^1$Institute of Space and Astronautical Science, Japan Aerospace Exploration Agency, 3-1-1 Yoshinodai, Sagamihara, Kanagawa 229-8510, Japan \\$^2$Hungarian Academy of Sciences, Research Group for Physical Geodesy and Geodynamics, Budapest P.O.Box 91., 1521 Hungary\\$^3$F\"OMI, Satellite Geodetic Observatory, Budapest P.O.Box 585, 1592 Hungary\\$^4$MPI f\"ur Radioastronomie, Auf dem H\"ugel 69, Bonn, 53121 Germany}   

\begin{abstract}

The blazar, S5 2007+77 shows intraday variability in cm wavelengths. Seven epochs of VSOP observations were carried out in 1998 and 1999 at 5\,GHz to look for the origin of the variability with the highest achievable angular resolution at cm wavelengths. Here the results of four epochs are analysed, which revealed $\sim 10$\,\% variations in polarized flux density.

\end{abstract}

\section{Introduction} 

S5 2007+77 is a blazar, showing pronounced variability across the electromagnetic spectrum at time scales ranging from years to minutes. It is also an Intraday Variable \citep[IDV, ][]{idv_discovery, idv_discovery2} source showing variations in total and in polarized intensity on time scales of 2 to 6 days at cm wavelengths \citep[e.g.][]{reduc_idv2,var2007}. IDV was also observed in the source at optical wavelengths, with amplitudes of $2$\,\% to $15$\,\% on a time scales of a few hours to a few days \citep[e.g.][]{opt}.

The IDV phenomenon is very common among the compact, flat-spectrum extragalactic radio sources. In many cases, IDV (at cm-wavelengths) can be explained and modelled as refractive interstellar scintillation \citep[RISS, e.g.][]{rickett1} caused by the ionized Interstellar Matter (ISM) of the Milky Way. However, there are sources, where several characteristics of the observed fast radio variations cannot be easily described via RISS models. For example, the correlation observed between the optical and radio intraday variations in S5 0716+714 and in 0954+658 \citep[e.g.][]{corr1,corr2} suggest a source intrinsic cause of IDV in these sources. 

Both source-extrinsic and source-intrinsic scenarios require very compact varying components (a few ten micro-arcseconds and a few micro-arcseconds, respectively). Therefore several IDV sources were proposed for observation with the Very Long Baseline Space Observatory Programme (VSOP) in the late 1990s. The aim of these observations was to look for the origin of the variability with the highest achievable angular resolution at cm wavelengths. These observations were planned so that one could distinguish from the measurements whether the short time scale variations can be attributed to the core or the jet of the particular radio source. The observations were performed at epochs separated by a few days and some weeks to allow to detect significant changes on short time scales. S5 2007+77 was among the proposed sources. (S5 0716+714 was also observed in this campaign, see the paper by Bach et al. in this proceeding.)

\section{Observation and Data Reduction}

In the space-VLBI observing campaign, S5 2007+77 was observed in seven epochs during 1998 and 1999. Here we present the four epochs which took place in 1999.679, 1999.685, 1999.753 and 1999.759. In all four epochs, the array consisted of the ten antennas of the Very Long Baseline Array (VLBA), the 100\,m radio telescope of the Max-Planck-Institut f\"ur Radioastronomie in Effelsberg (Germany) and the orbiting, 8\,m HALCA antenna of the VSOP. The observations were performed at 5\,GHz. The $(u,v)$ coverages were very similar in all four epochs, therefore the resulting maps from the different epochs can be reliably compared. 

The data reduction was done with the  National Radioastronomy Observatory (NRAO) Astronomical Image Processing System (AIPS) package following the usual procedures. HALCA was only able to detect left circular polarization (LCP), however space-VLBI polarimetry is still feasible. The HALCA LCP data can be correlated with the right circular polarization (RCP) data of the ground array. Because of this, only one of the cross polarization data is available, therefore instead of the standard imaging technique, complex imaging and complex deconvolution has to be applied to the $Q+iU$ dirty image \citep[see e.g.][]{pol1, pol2}.

In the gaps of the VLBI scans, the Effelsberg telescope worked as single-dish antenna and measured the (total and polarized) flux density of our target (S5 2007+77) and the two calibrators, J1800+782 and J1824+565, to ensure a proper calibration of the data. 

\section{Results}

\begin{figure}[!ht]
\plottwo{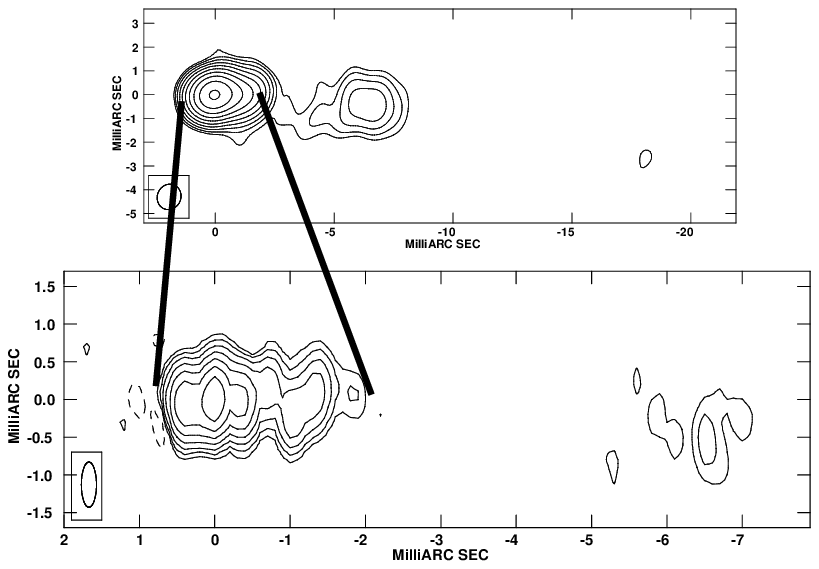}{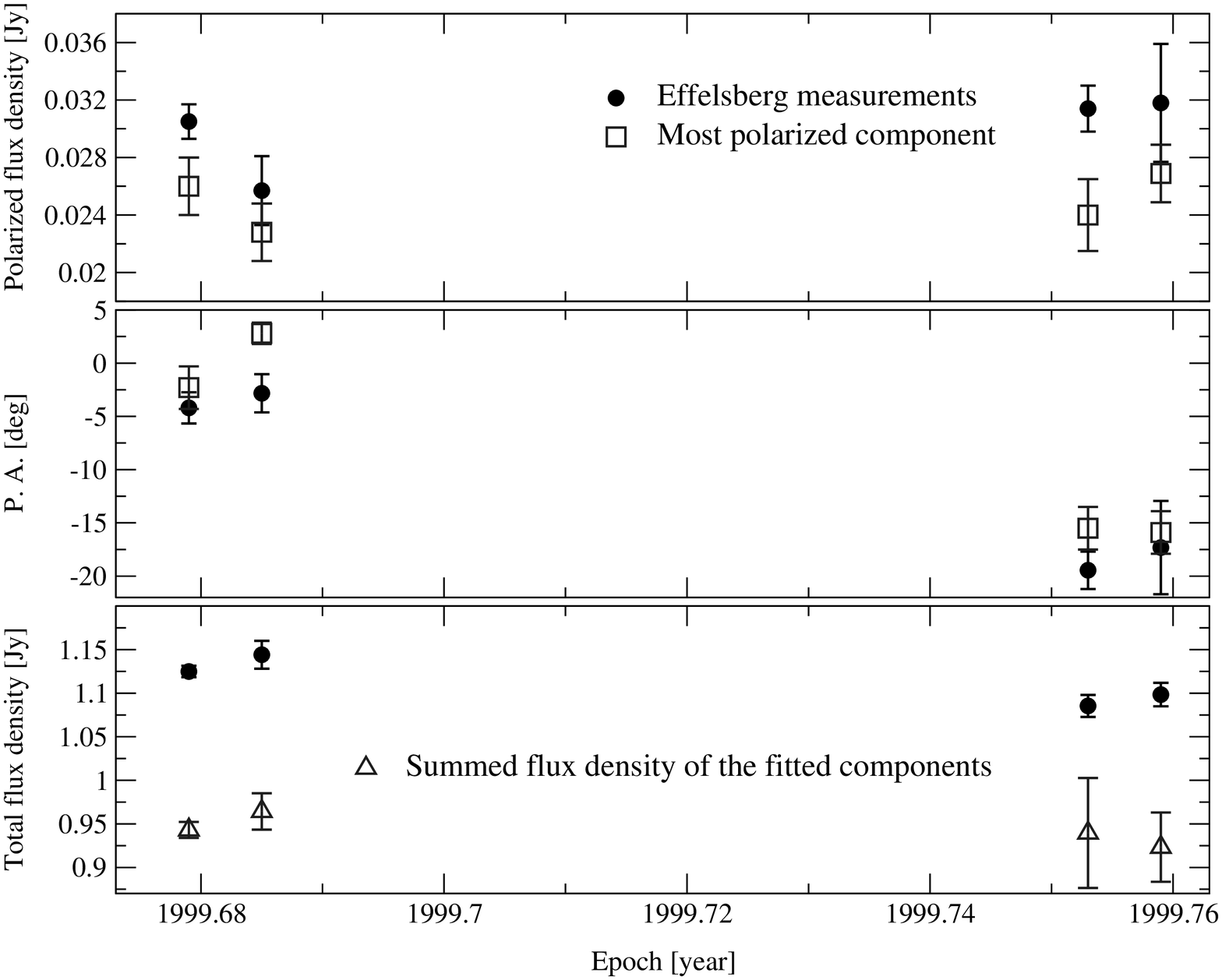}
\caption{\label{fig:maps}On the left hand side, contour maps of S5 2007+77 from the 1999.679 epoch. The top figure shows the map created using only the ground baselines. The bottom figure is created by using the space baselines as well. On the right hand side, the polarized flux density (top), the polarization angle (middle), and the total flux density of S5 2007+77 are displayed as measured by the Effelsberg telescope (filled circles) and the same for the most polarized feature in the VSOP images (open squares). In the bottom plot, open triangle represents the summed flux density of the fitted Gaussian model components to the VSOP data.}
\end{figure}

The ground-only total intensity data show a bright central region from which a $\sim 7$\,mas-long jet emerges westward, which ends in a knot-like feature (left side of Fig. \ref{fig:maps}). 
Using the much higher resolution provided by the space-baselines, the central bright region can be resolved into several sub-components. Eastwards from the brightest component a compact feature (possibly the core) can be seen at the higher resolution VSOP maps. We used the Difmap program to model fit the total intensity data with circular Gaussians.
The summed flux density of the fitted model components in every epoch is $\sim 85$\,\% of the flux density measured by the Effelsberg telescope. In total intensity only the brightest component showed a few percent of flux density variations.

\begin{figure}[!ht]
\plotone{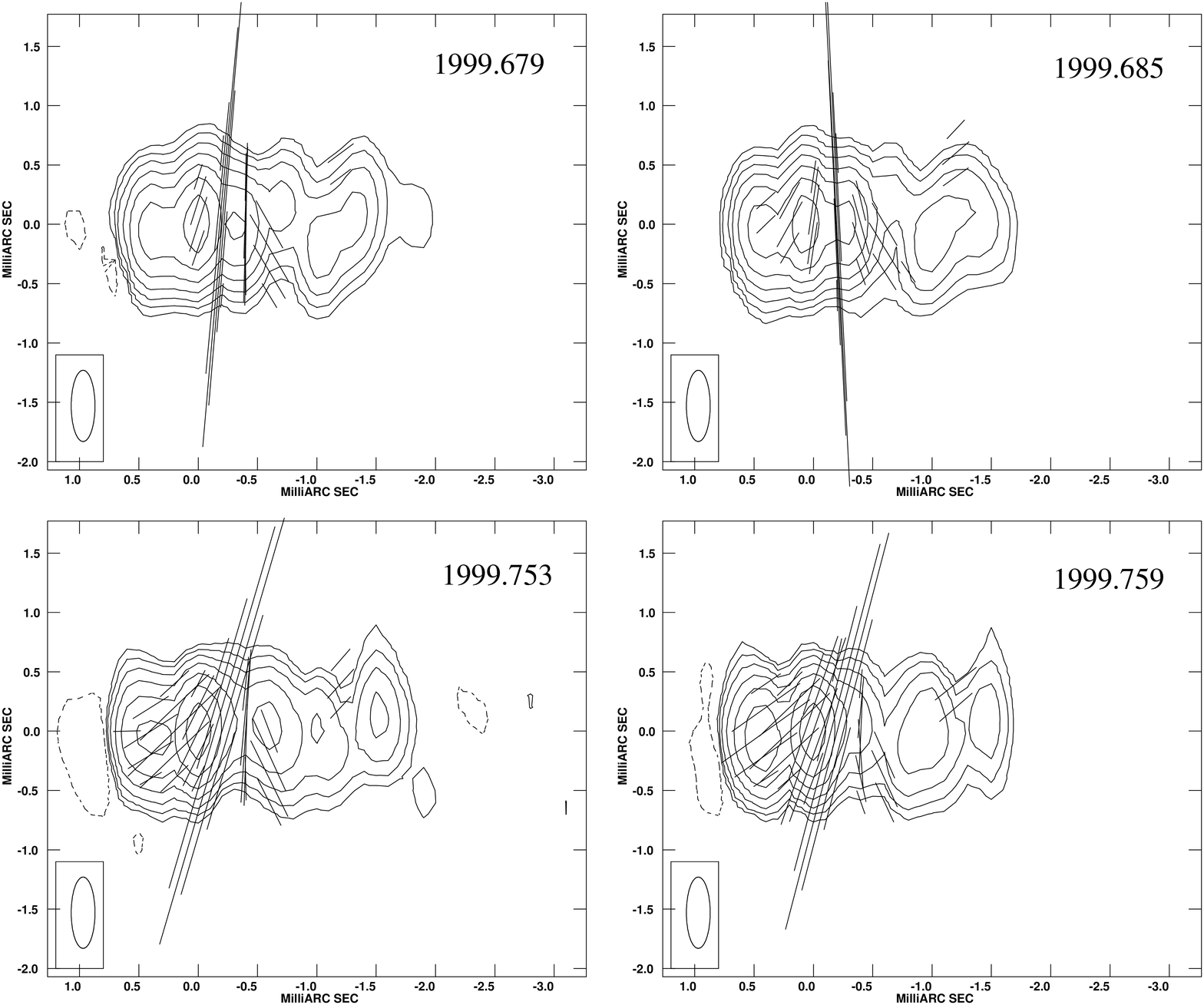}
\caption{\label{fig:pol}Space-VLBI maps of S5 2007+77. Contours show the total intensity; the lowest positive contours are 1\,\% of the peak intensity and the contour levels increase by a factor of 2. Peak intensities are $289$\,mJy/beam, $246$\,mJy/beam, $321$\,mJy/beam and $318$\,mJy/beam, for the four epochs respectively. The lines represent the direction of the EVPA, 1\,mas length corresponds to $5$\,mJy/beam polarized intensity.}
\end{figure}

In the high resolution polarization images (Fig. \ref{fig:pol}), three or four polarized components can be distinguished within the central $\sim 2$\,mas region. These features are slightly displaced from the features seen and model-fitted in the total intensity maps. The most polarized feature is not the brightest feature in total intensity, but rather an inter-knot region. The polarized flux density of the most polarized feature gives account for $\sim 75-85$\,\% of the polarized intensity measured by the Effelsberg telescope. The trend of variations of the polarized intensity and electric vector polarization angle of the most polarized feature also agrees well with the changes observed for the whole source with the Effelsberg telescope (see plot on the right hand side of Fig. \ref{fig:maps}). 

Between the first two epochs the polarized flux density decreased by $12$\,\% and the polarization angle of the most polarized feature changed $\sim 5$ degrees. This flux density increase using the brightness temperature calculation method of \cite{uwe} gives a variability brightness temperature of $\sim 10^{14}$\,K. On the longer time scale, between 1999.685 and 1999.753, the polarization angle changed by $\sim 12$\ degrees. While the polarized flux density of the most polarized feature somewhat decreased, the polarized flux density of the easternmost component (unpolarized in the first two epochs) increased to $\sim 10$\,mJy. Assuming that the component was in 1999.679 very close to the noise level $\sim 3$\,mJy and later, it brightened monotonically, we get a lower limit of $\sim10^{13}$\,K for the variability brightness temperature. These brightness temperature values imply a Doppler factors of $6$ and $3$ respectively (assuming $10^{12}$\,K inverse Compton brightness temperature limit). Using the additional VSOP epochs on S5 2007+77 a kinematic Doppler factor can be calculated and thus we will be able to check the agreement between the values obtained by the different methods.

\acknowledgements
This work made use of the VLBA, which is an instrument of the National Radio Astronomy Observatory, a facility of the National Science Foundation, operated under cooperative agreement by Associated Universities, Inc. This work is also based on observations with the 100 m radio telescope of the MPIfR (Max-Planck-Institut f\"ur Radioastronomie) at Effelsberg. We gratefully acknowledge the VSOP Project, which is led by the Japanese Institute of Space and Astronautical Science in cooperation with many organisations and radio telescopes around the world. We want to thank to C. Jin, Y. Hagiwara and B. Peng for their help in the data reduction.

\end{document}